\documentclass[%
 aip,
 amsmath,amssymb,
 reprint,%
]{revtex4-2}

\usepackage{graphicx}
\graphicspath{{figures/}}
\draft 
\usepackage{amsmath}
\usepackage{float}
\usepackage{siunitx}
\makeatletter
\let\newfloat\newfloat@ltx
\makeatother

\DeclareMathOperator{\sech}{sech}
\DeclareMathOperator{\asech}{asech}

\begin{document}


\title{Effects of parallel magnetic fields on sheaths near biased electrodes in a highly collisional Z-pinch plasma}



\author{C. R. Skolar}  
\email[]{chirag.skolar@njit.edu}
\affiliation{Center for Solar-Terrestrial Research, New Jersey Institute of Technology, Newark, NJ 07102, USA}

\author{B. Srinivasan}
\email[]{srinbhu@uw.edu}
\affiliation{William E. Boeing Department of Aeronautics and Astronautics, University of Washington, Seattle, WA 98195, USA}

\date{\today}

\begin{abstract}
		Sheath formation near biased electrodes in magnetic fields parallel to the wall is an understudied topic, especially within the context of Z-pinch fusion experiments.
		We perform 1X-2V Boltzmann-Poisson simulations of an axial cut at the pinch radius of a Z-pinch plasma between two biased electrodes with a magnetic field parallel to the wall.
		The collision frequencies are artificially increased to enhance thermalization of the plasma in the smaller simulation domain versus the actual experiment size;
		this increases the perpendicular mobility and partially de-magnetizes the ions resulting in non-monotonic sheath profiles with the potential increasing away from the wall to a peak before decaying.	
		A classical sheath forms within an electron gyroradius from the wall not due to the natural thermal motion of the electrons, but due to the magnetized electrons gyrating into the wall;
		therefore, the sheath structure does not significantly change with bias potential or between electrodes.
		With increasing bias potential, a current is induced perpendicular to the wall due to changes in ion flow, differing from unmagnetized cases where current is induced by changes in electron flow.
		The magnetic field acts as a high resistivity with the perpendicular current density being three orders of magnitude lower than unmagnetized theoretical predictions.
		There is, however, significant flow parallel to the wall from the force balance between the pressure tensor and Lorentz force.
		These parallel flows induce a parallel current density three orders of magnitude larger than the perpendicular current density.
\end{abstract}

	
\maketitle 

\section{Introduction}  
\label{s:intro}

Understanding the sheath structures that form near plasma facing surfaces is important for improving designs of magnetic confinement fusion walls,\cite{weynantsEdgeBiasingTokamaks1993}
electric thruster walls,\cite{wollenhauptOverviewThermalArcjet2018}
and Z-pinch electrodes.\cite{skolarContinuumKineticInvestigation2023,thompsonElectrodeDurabilityShearedflowstabilized2023,skolarGeneralKineticIoninduced2025}
Without the presence of a magnetic field, the interaction of a plasma with a solid wall results in a potential barrier near the wall\cite{robertsonSheathsLaboratorySpace2013}
whose thickness is on the order of a few Debye lengths, $\lambda_D=\sqrt{\epsilon_0 T_e/n e^2}$, 
where $\epsilon_0$ is the permittivity of free space,
$T_e$ is the electron temperature in energetic units,
$n$ is the density,
and $e$ is the elementary electric charge.
This is because the higher mobility electrons enter the wall first, charging it negative.
This repels further electrons and accelerates the ions to the wall resulting in a positive space charge near the wall.
In the literature, this phenomenon is typically called the sheath, plasma sheath, classical sheath, or ion sheath.

In the presence of a magnetic field with an oblique angle with respect to the wall, a magnetic presheath forms further from the wall due to the ion gyration.\cite{choduraPlasmaWallTransition1982,stangebyBohmChoduraPlasma1995}
Near the wall, however, a classical sheath still forms because there is a component of the magnetic field perpendicular to the wall allowing for plasma flow into the wall.
In the limiting case of a magnetic field entirely parallel to the wall, the ions and electrons will have significantly reduced mobility in the perpendicular direction, which is the direction toward the wall.
Therefore, due to their larger gyroradius, the ions enter the wall first, charging it positive.
This repels further ions and accelerates the electrons to the wall resulting in a negative space charge near the wall;\cite{krasheninnikovaScalingPlasmaSheath2010,krasheninnikovaEquilibriumPropertiesPlasma2010,moritzPlasmaSheathProperties2016,moritzPlasmawallTransitionLayers2018,liParticleSimulationMagnetized2018,liuTransitionIonElectron2025}
This phenomenon is often referred to as an inverse sheath, reverse sheath, or electron sheath.

In this work, the directions parallel and perpendicular are always with respect to the magnetic field.
Because the magnetic field is parallel to the wall, these directions can equivalently be considered with respect to the wall.

If the magnetic field is not sufficiently strong\cite{liuTransitionIonElectron2025} or if the collisional mean free path is sufficiently small,\cite{moritzPlasmawallTransitionLayers2018} the particles may be able to move across the field lines and a classical sheath will form instead.
Within the transition region between the inverse and classical sheaths for the collisional mean free path values, a non-monotonic potential profile may form that resembles a classical sheath with a potential barrier near the wall and then decaying further away.\cite{moritzPlasmawallTransitionLayers2018}

In addition, there are many applications where a plasma is in contact with biased electrodes, which may induce a variety of sheath behaviours\cite{baalrudInteractionBiasedElectrodes2020} and allow current to flow through the plasma.\cite{stangebySection26Potential2000,skolarContinuumKineticInvestigation2023,skolarGeneralKineticIoninduced2025}
The sheath structures near biased electrodes in the presence of magnetic fields have been understudied.\cite{baalrudInteractionBiasedElectrodes2020,liuTransitionIonElectron2025}
A priori, it is unclear if a bias potential may induce flow perpendicular to the parallel magnetic field to counter the magnetization of the plasma.
While this is an important fundamental plasma physics problem, this work is specifically motivated by Z-pinch plasma-material interactions.

Z-pinches are thermonuclear fusion reactor concepts that use biased electrodes to generate a current through a plasma.\cite{shumlakZpinchFusion2020}
The axial current induces an azimuthal magnetic field that confines and compresses the plasma increasing the density, temperature, and nuclear fusion reaction rate.
Understanding the current flow through the Z-pinch is important because the fusion energy yield scales as current to the eleventh power:\cite{shumlakZpinchFusion2020} $E_\text{fusion} \propto I^{11}$.
Previous work examining the sheath structure and current flow of Z-pinches has been unmagnetized.\cite{skolarContinuumKineticInvestigation2023,skolarGeneralKineticIoninduced2025}
This is valid at the radial center of the Z-pinch where the magnetic field is zero, but is not accurate away from the center where the magnetic field is non-zero.

The radial profile of the azimuthal magnetic field is found from a pressure balance between the plasma pressure and the magnetic pressure: $\nabla (p_i + p_e) = \mathbf{j}\times\mathbf{B}$, where
$p$ is the pressure for ions or electrons,
$\mathbf{j}$ is the current density,
and $\mathbf{B}$ is the magnetic field.
As a simplified approximation, we will consider the Bennett pinch profile\cite{bennettMagneticallySelfFocussingStreams1934} assuming a radially isothermal plasma with stationary ions and current driven entirely by uniform electron flow.
While these assumptions are not valid in a Z-pinch (the plasma is not isothermal\cite{geykoGyrokineticExtendedMHDSimulations2021} and the current is carried by a combination of the ions and electrons\cite{skolarContinuumKineticInvestigation2023,skolarGeneralKineticIoninduced2025}), they provide an analytical approximation that can easily be modeled.
Future work should consider more realistic radial equilibria based on theory\cite{crewsKadomtsevPinchRevisited2024} or experiment.\cite{geykoGyrokineticExtendedMHDSimulations2021}
Given these assumptions, the resulting density and magnetic field from the pressure balance are\cite{hainesReviewDenseZpinch2011,allenBennettPinchNonrelativistic2018}
\begin{align}
	n &= \frac{n_0}{\big(1 + b n_0 r^2\big)^2}
	\label{eq:n_Bennett} \\
	B &= \frac{\mu_0 e u n_0 r}{2 \big( 1 + b n_0 r^2 \big)},
	\label{eq:B_Bennett}
\end{align}
where $r$ is the radius,
$n_0$ is the density at the center of the Z-pinch, 
$\mu_0$ is the vacuum permeability,
$u$ is the electron velocity,
and
\begin{equation}
	b = \frac{\mu_0 e^2 u^2}{8 \big(T_e + T_i \big)}
	\label{eq:Bennett_param}
\end{equation}
is a length scale called the Bennett parameter with $T$ as the ion or electron temperature.

\begin{figure}[!htb]
	\centering
	\includegraphics[width=\linewidth]{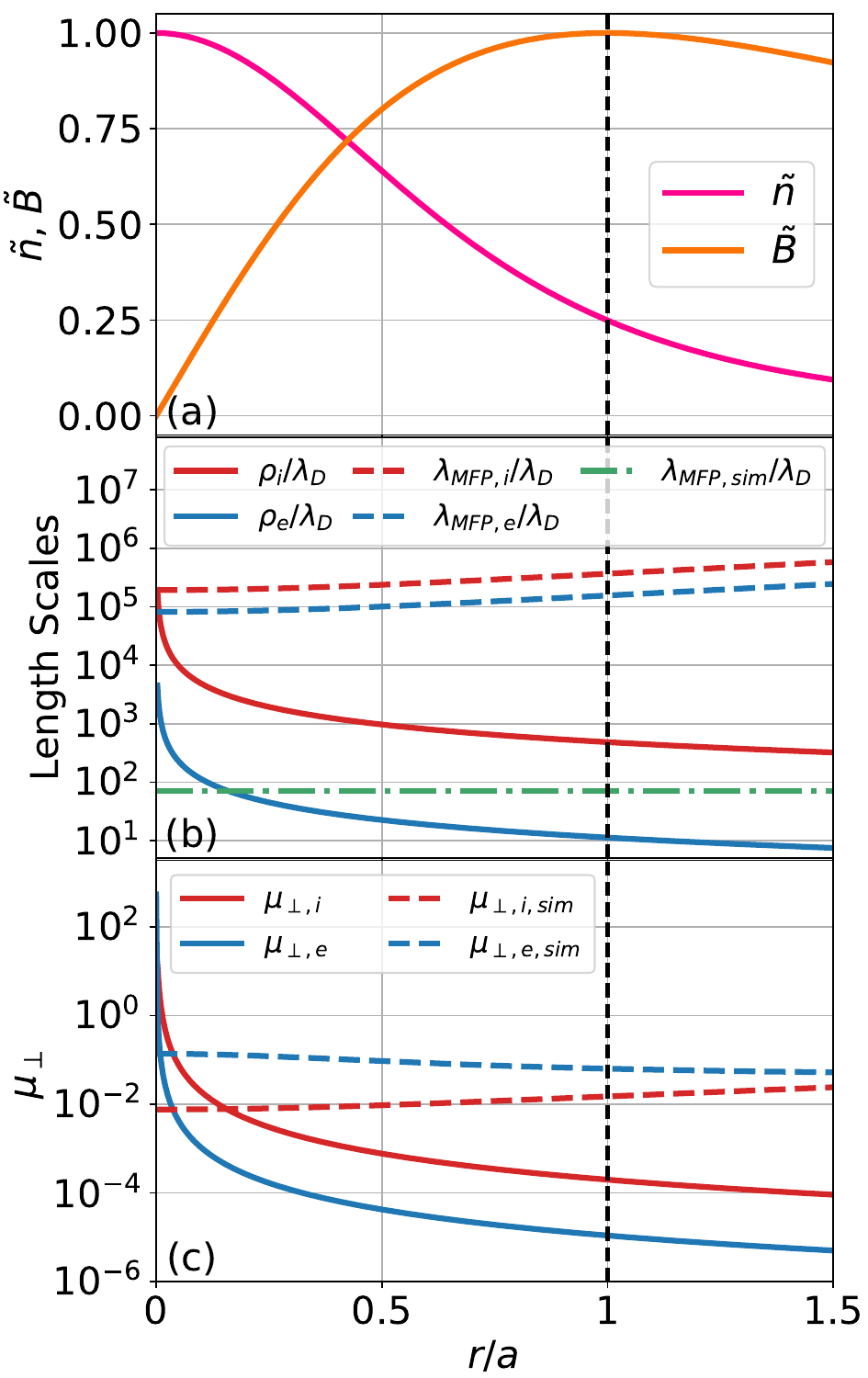}
	\caption{Panel (a) shows the radial normalized Bennett pinch profiles of the density (pink) and magnetic field (orange) based on Eqs.~\ref{eq:n_Bennett} and \ref{eq:B_Bennett}, respectively.
		Panel (b) shows how important length scales, normalized by the Debye length, vary with radius:
		ion gyroradius (solid red line),
		electron gyroradius (solid blue line),
		ion mean free path (dashed red line),
		electron mean free path (dashed blue line),
		and artificial simulation mean free path (dash-dot green line).
		The ion and electron mean free paths are based on the total collision frequencies in Eqs.~\ref{eq:nu_e} and \ref{eq:nu_i}, respectively.
		The artificial simulation mean free path is the same for ion and electron self-collisions at $50\sqrt{2}\lambda_D$.
		Panel (c) shows the perpendicular mobility of the ions (red) and electrons (blue) based on realistic Z-pinch parameters\cite{zhangSustainedNeutronProduction2019} (solid line) or with artificially increased collisions (dashed line).
		The vertical black dashed line corresponds to the pinch radius, which is the radius at which the simulations in this paper are run.}
	\label{fig:bennett}
\end{figure}

Figure~\ref{fig:bennett} presents several parameters as a function of radius within ($r/a\leq1$) and outside ($r/a>1$) of a Z-pinch.
The parameter $a$ is the pinch radius and considered to be the edge of the Z-pinch.
Figure~\ref{fig:bennett}(a) shows the density and magnetic field profiles, normalized by their maximum values, based on Eq.~\ref{eq:n_Bennett} and \ref{eq:B_Bennett}, respectively.
We use $n_0=\SI{1.1e23}{\meter^{-3}}$, $T_i=T_e=T_0=\SI{2}{\kilo\electronvolt}$, $I=\SI{200}{\kilo\ampere}$, and $a=\SI{3}{\milli\meter}$ based on experimental data from the Fusion Z-pinch Experiment (FuZE).\cite{zhangSustainedNeutronProduction2019}
The current is related to the axial electron velocity by $u=I/en_0\pi a^2$.
The density has a peak at $r=0$ and decays with increasing radius.
The magnetic field is zero at $r=0$ and increases to a maximum value at the pinch radius, $r/a=1$;
then, it decays as the radius increases further.

The changing density and magnetic field affect key length scales in the Z-pinch plasma, as shown in Fig.~\ref{fig:bennett}(b).
Assuming an isothermal plasma, the Debye length increases with radius because the density decreases with radius.
Based on the combination of the change in magnetic field strength and density, the ion and electron gyroradii, normalized to the Debye length, decrease with radius.
We define the gyroradius of species $\alpha$ to be $\rho_\alpha = m_\alpha v_{th,\alpha} / e B$,
where $v_{th,\alpha}=\sqrt{2T_\alpha/m_\alpha}$ is the thermal velocity.
Due to the difference in mass ratio, the electron gyroradius is smaller than the ion gyroradius.
From a computational perspective, the simulation needs to resolve both the ion gyroradius (because it allows for possible formation of the magnetic presheath) and the Debye length (which is the length scale that defines classical sheath behavior).
Because of the large disparity between these two values, we choose to use the values at the pinch radius ($r/a=1$) where the ratio of the ion gyroradius to the Debye length is 486.
This allows for a reasonably sized and computationally tractable simulation domain of $\pm 2048\lambda_D$.

Ref.~\onlinecite{moritzPlasmawallTransitionLayers2018} show that the sheath structure that develops is related to the size of the collisional mean free path compared to the ion and electron gyroradii.
In our work, we will only consider the combination of collisions between ions and electrons (no inclusion of neutral particles).
Based on the densities from Eq.~\ref{eq:n_Bennett}, the collision frequency variations with radius can be found using
\begin{align}
	\nu_{ee} &= \frac{n e^4 }{2 \pi \epsilon_0^2 m_e^2 v_{th,e}^3} \ln \Big( \frac{\Lambda}{3} \Big) \label{eq:nu_ee} \\
	\nu_{ei} &= \nu_{ee}\sqrt{2} \label{eq:nu_ei}\\
	\nu_{ie} &= \frac{m_e}{m_i} \nu_{ee} \label{eq:nu_ie}\\
	\nu_{ii} &= \nu_{ee}\sqrt{\frac{m_e}{m_i}} \label{eq:nu_ii} \\
	\nu_e &= \nu_{ee} + \nu_{ei} \label{eq:nu_e} \\
	\nu_i &= \nu_{ii} + \nu_{ie}, \label{eq:nu_i}
\end{align}
where $\nu_{\alpha\beta}$ is the collision frequency of particle $\alpha$ colliding with particle $\beta$ and $\Lambda=12\pi n \lambda_D^3$ is the number of particles in the Debye sphere.
Equations~\ref{eq:nu_e} and \ref{eq:nu_i} are the total collision frequencies of the electrons and ions, respectively.
The mean free path is related to the collision frequency and the thermal velocity through $\lambda_{MFP,\alpha} = v_{th,\alpha}/\nu_\alpha$.
Figure~\ref{fig:bennett}(b) shows that both species' mean free paths (dashed lines) are always larger than the electron gyroradius (solid blue line).
Except for an incredibly small region near $r=0$, the mean free paths are also larger than the ion gyroradius (solid red line).
This suggests that the plasma would be highly magnetized as there are very few collisions within one gyro-period.
This is shown by examining the perpendicular mobility, in Fig.~\ref{fig:bennett}(c), defined as
\begin{equation}
	\mu_{\perp, \alpha} = \frac{\mu_\alpha}{ 1 + \Omega_{c\alpha}^2/\nu_\alpha^2}, \label{eq:perp_mobility}
\end{equation}
where $\Omega_{c\alpha}=eB/m_\alpha$ is the gyrofrequency and $\mu_\alpha=e/m_\alpha \nu_\alpha$ is the mobility.
Figure~\ref{fig:bennett}(c) shows that the perpendicular mobility (solid lines) decreases with radius.

While the simulation domain is chosen to be $\pm 2048\lambda_D$, the actual experiment size\cite{zhangSustainedNeutronProduction2019} is \SI{50}{\centi\meter} or about $\num{250000}\lambda_D$ at $r/a=1$.
To account for this large difference and ensure thermalization of the plasma in the center of the domain, we artificially increase the collision frequency such that the ion and electron self-species collisions have mean free paths of $50 \sqrt{2} \lambda_D$.
Based on this increased collision frequency, the simulation mean free path (green dash-dot line) is larger than the electron gyroradius but smaller than the ion gyroradius at $r/a=1$, as shown in Fig.~\ref{fig:bennett}(b).
This also significantly increases the perpendicular mobility, as shown in Fig.~\ref{fig:bennett}(c) (dashed lines).
Therefore, the parameter regime with the ion mean free path smaller than the ion gyroradius and an increased perpendicular mobility used in this work can be considered as an extreme case for highly collisional Z-pinch plasmas.
The results still shed light on the magnetized sheath behavior and provide important context for what will likely occur in lower and more realistic collision regimes.

In this work, we use the continuum kinetic approach to study the effect of a parallel magnetic field on the sheath and current flow of a plasma between biased electrodes at the pinch radius ($r=a$) of a Z-pinch plasma.
Section~\ref{s:sim_setup} describes the computational methods and simulation setup.
Section~\ref{s:results} presents the results of how the magnetic field and potential bias affect the sheath structure and current flow.
Section~\ref{s:conclusions} provides the summary and conclusions of the paper.

\section{Simulation Setup}
\label{s:sim_setup}

We model an electrostatic proton-electron plasma that is doubly bounded in the $z$ direction by electrodes with an electric potential bias between them.
There is a static uniform background magnetic field in the $\theta$ direction.
Spatially, we are modeling an axial cut of the Z-pinch plasma at a radius of $r=a=\SI{3}{\milli\meter}$, which is the pinch radius of FuZE.\cite{zhangSustainedNeutronProduction2019}

We evolve the distribution function using the 1X-2V, or one spatial ($z$) and two velocity ($v_z$ and $v_r$) dimensions, Boltzmann equation,
\begin{multline}
	\frac{\partial f_\alpha}{\partial t} + \nabla_\textbf{x} \cdot \big( f_\alpha \mathbf{v} \big)
	+ \frac{q_\alpha}{m_\alpha} \nabla_\mathbf{v} \cdot \bigg[ f_\alpha \Big( \mathbf{E} + \mathbf{v}\times\mathbf{B}    \Big) \bigg] 
	\\= \sum \bigg( \frac{\partial f_\alpha}{\partial t} \bigg)_c + S_\alpha,
	\label{eq:boltzmann}
\end{multline}
where $f$ is the distribution function,
$q$ is the electric charge,
and $m$ is the mass of species $\alpha$ (ions or electrons).

The summation term is the Dougherty or Lenard-Bernstein collision operator\cite{doughertyModelFokkerPlanckEquation1964,francisquezConservativeDiscontinuousGalerkin2020,hakimConservativeDiscontinuousGalerkin2020} which includes the effects of inter- and intra-species collisions through the collision frequencies from Eqs.~\ref{eq:nu_ee}-\ref{eq:nu_ii}.
We artificially increase the electron-electron collision frequency, $\nu_{ee}$, across the entire domain to have a mean free path of $50\sqrt{2}\lambda_D$ to allow for greater thermalization in the smaller domain compared to the full experiment.
This also has the effect of artificially increasing the perpendicular mobility, as shown in Fig.~\ref{fig:bennett}(c).

A source term, $S_\alpha$, is included to maintain particle conservation within the domain.\cite{liBohmCriterionPlasma2022,liTransportPhysicsDependence2022}
The shape of the source term is chosen to be largest in the center of the domain, tend to zero towards the walls, and be $C^\infty$ continuous.
It is defined as
\begin{equation}
	S_\alpha = c_1 \sech^2 \big( c_2 z \big) \Gamma_{i,out} f_{0,\alpha}(v_z, v_r),
\end{equation}
where $\Gamma_{i,out}$ is the ion particle flux leaving the domain,
$f_{0,\alpha}(v_z, v_r)$ is a 2V normalized Maxwellian with the initial temperature and unity density of species $\alpha$,
and the coefficients $c_1$ and $c_2$ are
\begin{align}  
	c_1 &= \frac{c_2}{2 \tanh \big( 2048\lambda_D c_2  \big)} \label{eq:c1} \\
	c_2 &= \frac{\asech \big(\sqrt{0.01}\big)}{1248\lambda_D}. \label{eq:c2}
\end{align}

The electric field is related to the electric potential through $\mathbf{E} = - \nabla \phi$.
The electric potential is found using the Poisson equation,
\begin{equation}
	\nabla^2 \phi = - \frac{e (n_i - n_e)}{\epsilon_0},
	\label{eq:poisson}
\end{equation}
where $n_i$ and $n_e$ are the ion and electron number densities, respectively, which are found by taking the zeroth moment of their respective distribution functions.

The discontinuous Galerkin code \verb|Gkeyll|\cite{junoDiscontinuousGalerkinAlgorithms2018,hakimAliasFreeMatrixFreeQuadratureFree2020} is used to solve Eqs.~\ref{eq:boltzmann} and \ref{eq:poisson}. We use an orthonormal modal serendipity basis with second order polynomials for the discretization.
Equation~\ref{eq:boltzmann} is evolved in time using a three stage third order strong stability preserving Runge-Kutta method.\cite{gottliebHighOrderStrong2005}
The Poisson equation, Eq.~\ref{eq:poisson}, is solved via direct matrix inversion.
Variations of this numerical model have been used for a variety of plasma-material interactions studies.\cite{cagasContinuumKineticMultifluid2017,cagasBoundaryValueReservoir2021a,skolarContinuumKineticInvestigation2023,bradshawEnergydependentImplementationSecondary2024,bradshawEffectsOxidationImpurities2025,skolarGeneralKineticIoninduced2025}

Zero-flux boundary conditions are used for the velocity space ($v_z$ and $v_r$).
For configuration space ($z$), absorbing wall boundary conditions are used.
Dirichlet boundary conditions are used for the electric potential with the left wall set to an applied bias potential, $\phi_b$, and the right wall set to zero.
Therefore, the left and right walls are defined as the anode and cathode, respectively.

The plasma is initialized as a uniform Maxwellian with a temperature of \SI{2}{\kilo\electronvolt} and a density based on Eq.~\ref{eq:n_Bennett} or Fig.~\ref{fig:bennett}(a) at $r=a=\SI{3}{\milli\meter}$.
This corresponds to an initial uniform density of \SI{2.741e22}{\meter^{-3}}.
The static background magnetic field in the $\theta$ direction is \SI{6.655}{\tesla} based on Eq.~\ref{eq:B_Bennett} or Fig.~\ref{fig:bennett}(a) at $r=a$.
We run three separate cases with bias potentials of 0, 5, and \SI{10}{\kilo\volt}.

The configuration space domain size is set to be $\pm 2048\lambda_D$ which fully resolves both the ion and electron gyroradii, which are $486\lambda_D$ and $11.3\lambda_D$, respectively.
In addition, the Debye length scales are resolved with a configuration space resolution of two cells per Debye length or 8192 cells total.
The $v_z$ and $v_r$ direction velocity spaces for the electrons span $\pm 2\sqrt{2} v_{th,e}$ with 16 cells and $\pm 2\sqrt{2} v_{th,e}$ with 8 cells, respectively.
The $v_z$ and $v_r$ direction velocity spaces for the ions span $\pm 3\sqrt{2} v_{th,e}$ with 16 cells and $\pm 2\sqrt{2} v_{th,e}$ with 8 cells, respectively.
The additional domain in $v_z$ for the ions is to account for the ion acceleration that typically occurs near the wall in a plasma sheath.
The coarse resolution in $v_z$ has been found to provide converged results.\cite{cagasContinuumKineticMultifluid2017}
Furthermore, the even coarser resolution in $v_r$ provides an adequate description because the distribution functions are generally Maxwellian in $v_r$.

\section{Results} \label{s:results}
All the simulations are run until a steady state is achieved.
The data for the 0, 5, and \SI{10}{\kilo\volt} cases are taken at times of \num{2.64}, \num{2.70}, and \num{2.35} ion gyroperiods, respectively.
An ion gyroperiod is defined as $2\pi/\Omega_{ci}$.

\begin{figure}[!htb]
	\centering
	\includegraphics[width=\linewidth]{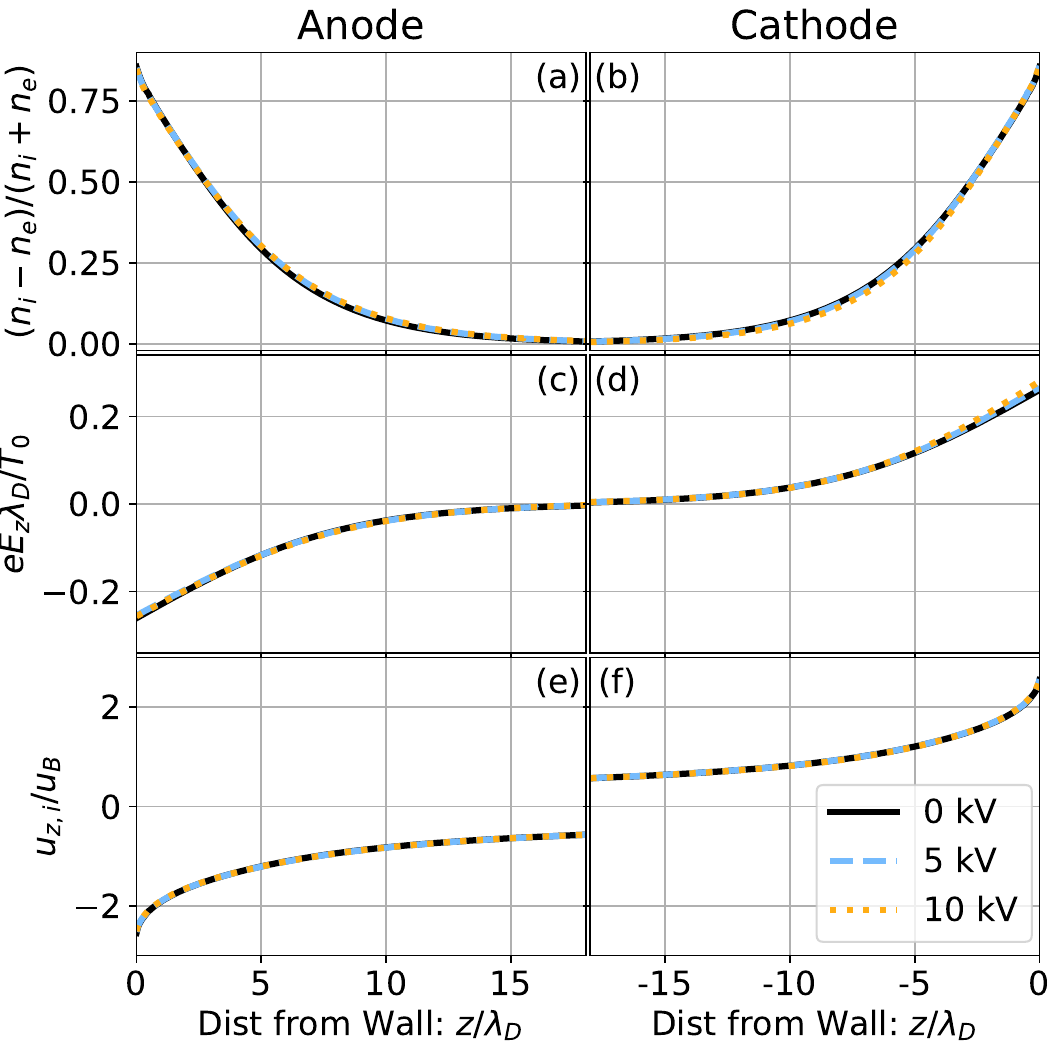}
	\caption{Plots of the fractional charge density (a-b), normalized axial electric field (c-d), and normalized axial ion drift velocity (e-f) at the anode (a,c,e) and cathode (b,d,f) for bias potentials of 0 (solid black line), 5 (dashed blue line) and \SI{10}{\kilo\volt} (dotted yellow line). Minimal changes in sheath behavior are observed between these cases and between the electrodes.} 
	\label{fig:sheath_profile}
\end{figure}

Figure~\ref{fig:sheath_profile} shows the plasma sheath structure near the anode (a,c,e) and cathode (b,d,f) for all three bias potential cases.
Panels (a-b) show the fractional charge density, which is defined as $(n_i-n_e)/(n_i+n_e)$.
At both electrodes, the fractional charge density shows that there exists a positive space charge near the wall with minimal to no change as the bias potential increases.

Panels (c-d) show the axial electric field normalized by the Debye length and the initial temperature.
As with the density, the electric field generally stays the same at both electrodes with only minor visible changes at the cathode.

Panels (e-f) show the axial ion velocity normalized by the local Bohm speed which is defined as 
$u_B = \sqrt{ \gamma (T_{z,e} + T_{z,i})/m_i }$ where $\gamma=2$ based on the two degrees of freedom in 1X-2V simulations.
Beyond sign (due to the walls facing different directions), there is no difference in the axial ion velocity between electrodes or between bias potentials.
This suggests that the sheath entrance does not change with bias potential.

For all of these profiles, it is clear that there is minimal to no change in the plasma sheath between electrodes and between bias potentials.
This is counter to predictions from simplified unmagnetized theory\cite{stangebySection26Potential2000} and previous unmagnetized simulation studies.\cite{skolarContinuumKineticInvestigation2023,skolarGeneralKineticIoninduced2025}
These works noted substantial changes in the densities, electric field, ion velocity, and sheath length.
Therefore, the inclusion of a parallel magnetic field acts to mitigate the effects of a bias potential on the plasma sheath.

\begin{figure*}[!htb] 
	\includegraphics[width=\linewidth]{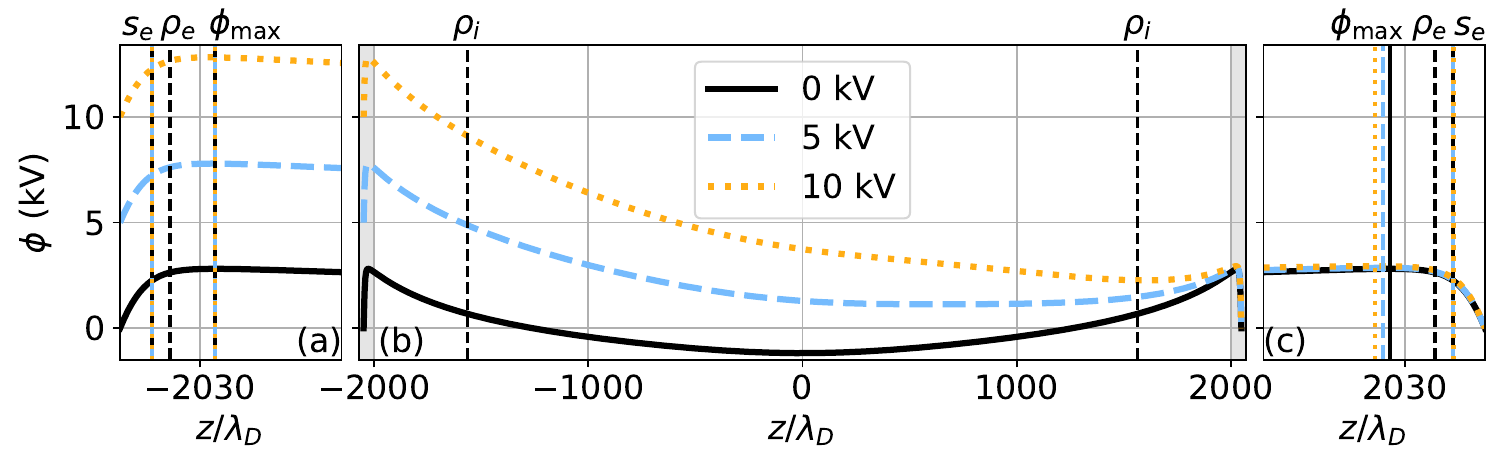}
	\caption{Plot of the potential profile (b) for bias potentials of 0 (solid black line), 5 (dashed blue line), and \SI{10}{\kilo\volt} (dotted yellow line).
	The lightly shaded regions in Panel (b) are presented in expanded scales in Panels (a) and (c).
	The vertical dashed lines correspond to important length scales, annotated above the plots.}
	\label{fig:potential}
\end{figure*}

This is further evidenced by examining the electric potential profile throughout the entire domain, as shown in Fig.~\ref{fig:potential}(b).
For the \SI{0}{\kilo\volt} case, the electric potential increases away from the wall up to a peak value.
Then, the potential decreases until the center of the domain where it reaches a minimum value below the wall potentials.
Because there is no bias potential in this case, the potential profile is symmetric about $z=0$.
With the inclusion of a bias potential, the potential profile is no longer symmetric.
Both the 5 and \SI{10}{\kilo\volt} cases behave similarly, just with higher magnitudes for the \SI{10}{\kilo\volt} case.
Starting from the cathode (right), the potential is at \SI{0}{\kilo\volt}, as defined by the boundary condition.
Moving into the the domain toward the left, the potential increases up to a peak value. 
Then the potential decreases slightly before increasing up to another peak near the anode (right).
The potential then drops back down to the specified boundary condition on the left.
This non-monotonicity in the potential profile is not observed in unmagnetized sheaths.\cite{stangebySection26Potential2000,skolarContinuumKineticInvestigation2023,skolarGeneralKineticIoninduced2025}

The non-monotonicity has, however, been observed in collisional magnetized sheaths and can be explained by the effect of differences in charge density on the ambipolar electric field.\cite{moritzPlasmawallTransitionLayers2018}
While Ref.~\onlinecite{moritzPlasmawallTransitionLayers2018} considered only collisions with neutrals, the general idea is the same for our simulations without neutrals that only have collisions with ions or electrons.
When the mean free path is smaller than the gyroradius, the particles will generally undergo more collisions on average per gyro-orbit allowing for some de-magnetization of the plasma.
In our case, the chosen mean free path (for both ions and electrons) of $50\sqrt{2}\lambda_D$ sets the length scale ordering to $\rho_e < \lambda_{MFP} < \rho_i$, as seen in Fig.~\ref{fig:bennett}(b) at $r/a=1$.
The electrons are heavily magnetized because the electron gyroradius ($11.3\lambda_D$) is smaller than the mean free path.
The ions are less magnetized because the ion gyroradius ($486\lambda_D$) is larger than the mean free path.

Figures ~\ref{fig:potential}(a,c) present expanded scales of the sheath regions near the electrodes, which are shaded in gray in Fig.~\ref{fig:potential}(b).
At these smaller scales, we can see that the electric potential changes rapidly near the wall, but slowly further away, resulting in a plateau upon reaching the peak value.
This is further shown in Figs.~\ref{fig:sheath_profile}(c-d) where the electric field quickly decays close to zero far from the wall showing how slowly the potential changes.
Overlaid onto Fig.~\ref{fig:potential} are several important length scales to obtain an understanding of the physical mechanism causing the non-monotonic potential profile.
As shown in Figs.~\ref{fig:sheath_profile}(e-f), the sheath entrance ($s_e$), defined as where $|u_{z,i}|/u_B=1$, does not significantly change with bias potential at either electrode.
Compared to where the potential reaches a plateau, the sheath entrance is closer to the wall.
While this may help explain the classical sheath behavior directly near the wall, it does not explain the decrease in potential further away from the wall.

This is explained by looking at the electron gyroradius, which occurs around where the potential reaches a plateau.
The electron gyroradius is an average quantity and the temperature also varies within the sheath region; 
therefore, it is not expected to be a hard delineation point, hence why the location of peak potential, $\phi_{\max}$, is still slightly further into the domain.
Within the electron gyroradius, the electrons gyrate about the magnetic field and generally strike the wall before the ions, charging the wall negative as in a classical sheath.
This results in the classical sheath behavior directly near the wall with a potential increase that accelerates ions to the wall.
Furthermore, because the sheath is caused by the electron gyro-motion and not natural flow of electrons into the wall, bias potentials do not significantly change the behavior, explaining Fig.~\ref{fig:sheath_profile}, with the sheath being confined within the electron gyroradius.

The decrease in the potential further from the wall is explained by the ions gyrating into the wall from further into the domain. 
Figs.~\ref{fig:sheath_profile}(a-b) do not show a visible difference in the fractional charge density far from the wall.
However, by the decrease in the potential toward the center of the domain, it is clear that a charge difference develops with more ions than electrons due to ion gyration.
The reason the ion gyration does not dominate and form an inverse sheath from ions impacting the wall before electrons is because of the increased ion collision frequency;
the ion mean free path ($50\sqrt{2}\lambda_D$) is smaller than the ion gyroradius ($486\lambda_D$) decreasing the amount of ions that gyrate into the wall.

Therefore, the non-monotonicity of the potential profile is explained by the electrons inducing a classical sheath near the wall due to their gyroradius being smaller than the mean free path and the ions reducing the potential further from the wall, but at a slower rate, due to their gyroradius being larger than the mean free path.

It is important to note that we artificially increased the collisions in our simulations.
If we consider instead the realistic length scales from Fig.~\ref{fig:bennett}, the ordering at $r=a$ changes to $\rho_e < \rho_i < \lambda_{MFP, e} < \lambda_{MFP, i}$.
Based on this ordering, neither the electrons nor the ions are sufficiently collisional and therefore the non-monotonic potential profile would not be induced.
Instead, Ref.~\onlinecite{moritzPlasmawallTransitionLayers2018} suggests an inverse sheath would form instead as there would not be sufficient perpendicular mobility.
Future work should examine these lower collisionality regimes.

\begin{figure}[!htb]  
	\centering
	\includegraphics[width=\linewidth]{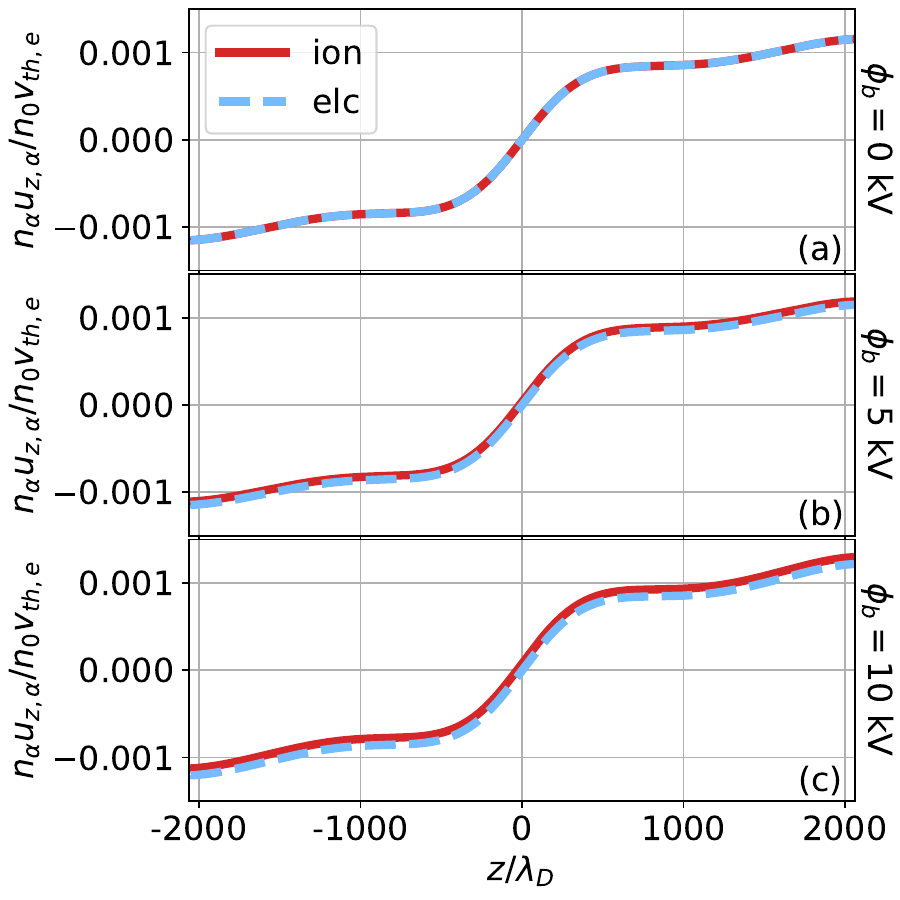}
	\caption{Plots of the axial ion (solid red line) and electron (dashed blue line) particle fluxes for bias potentials of 0 (a), 5 (b), and 10 (b) \si{\kilo\volt}. 
	Both particle fluxes slightly increase in magnitude with increasing bias potential.
	The ion particle flux becomes more positive with increasing bias potential resulting in a net axial current toward the cathode.}
	\label{fig:nu_x}
\end{figure}

Despite the sheath behavior generally not changing with bias potential, there is clearly an electric field, albeit small, within the center region of the domain for the 5 and \SI{10}{\kilo\volt} cases as shown by the steady leftward increase in potential in Fig.~\ref{fig:potential}(b).
Therefore, a current is expected to flow between the electrodes, as has been shown in previous unmagnetized works.\cite{stangebySection26Potential2000,skolarContinuumKineticInvestigation2023,skolarGeneralKineticIoninduced2025}
Figure~\ref{fig:nu_x} shows the axial ion (solid red line) and electron (dashed blue line) particle fluxes, normalized by the initial density and electron thermal velocity, for the 0 (a), 5 (b), and 10 (c) \si{\kilo\volt} cases.

For the \SI{0}{\kilo\volt} case, the axial ion and electron particle fluxes are the same, resulting in no net current, as should be expected without a bias potential.
At a bias potential of \SI{5}{\kilo\volt}, the magnitude of the axial particle fluxes for both species becomes slightly larger and the axial ion particle flux shifts to being greater than that of the electrons throughout the entire domain.
This results in a small net current driven by the additional ion flow.
These effects are seen in greater magnitudes for the \SI{10}{\kilo\volt} case.
The current being driven by changes in the ion flow is counter to what is typically seen in unmagnetized biased sheaths where the electron flow changes substantially to induce significant current.\cite{stangebySection26Potential2000,skolarContinuumKineticInvestigation2023,skolarGeneralKineticIoninduced2025}
With the magnetic field, the electrons are heavily magnetized and their axial flow does not significantly change with bias potential.
Therefore, because the ions are less magnetized, they drive the current.
The current in our simulations, however, is much smaller than that predicted by unmagnetized sheath theory. 
For high bias potentials, such as \SI{10}{\kilo\volt}, unmagnetized sheath simulations found that the current reached a saturation limit\cite{skolarContinuumKineticInvestigation2023,skolarGeneralKineticIoninduced2025} which matched the theoretical prediction of $e n_0 c_s/2$ where $c_s$ is the ion sound speed.\cite{stangebySection26Potential2000}
From this theory, the current density at the pinch radius would be on the order of \SI{e9}{\ampere\meter^{-2}} whereas it is only on the order of \SI{e6}{\ampere\meter^{-2}} in our simulations, which is a significant reduction.
This reduction is further exacerbated because these simulations have artificially increased the perpendicular mobility, as shown in Fig.~\ref{fig:bennett}(c);
therefore, for less mobile realistic Z-pinch plasmas, the current will likely be even lower.
The parallel magnetic field acts as a large effective resistivity preventing significant current from flowing through the plasma at the pinch radius.

\begin{figure}[!htb]
	\centering
	\includegraphics[width=\linewidth]{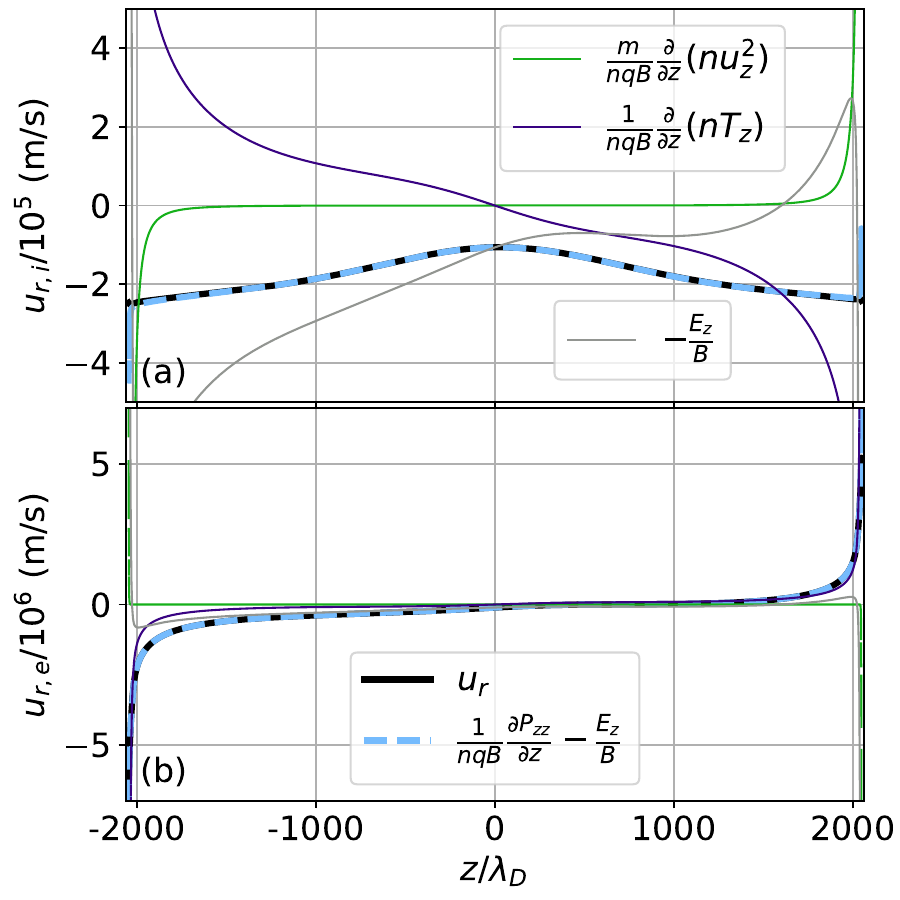}
	\caption{Plots of the radial ion (a) and electron (b) velocities for the \SI{10}{\kilo\volt} case. 
		The radial velocities (solid black line) occur due to a force balance between the pressure tensor and Lorentz force (dashed blue line), as described by Eq.~\ref{eq:uy_force_balance}.
		Also plotted are the individual drifts:
		inertial drift (solid green line),
		diamagnetic drift (solid purple line),
		and $\mathbf{E}\times\mathbf{B}$ drift (solid gray line).
	}
	\label{fig:u_y_force}
\end{figure}

At equilibrium, the forces in the axial direction induce velocities in the radial direction through the Lorentz force, as shown in Fig.~\ref{fig:u_y_force}.
The steady state momentum equation is
\begin{equation}
	\nabla \cdot \mathbf{P_\alpha} = n_\alpha q_\alpha \Big( \mathbf{E} + \mathbf{u}_\alpha \times \mathbf{B} \Big)
	\label{eq:mtm}
\end{equation}
where the elements, $P_{ij,\alpha}$, of the pressure tensor, $\mathbf{P}_\alpha$ are defined as
\begin{equation}
	P_{ij,\alpha} = m_\alpha \int_{-\infty}^\infty v_i v_j f_\alpha d\mathbf{v},
\end{equation}
or the second moment of the distribution function.
In our 1X-2V setup, we only need to consider the $z$ component of the electric field, the $z$ and $r$ components of the drift velocities, the $\theta$ component of the magnetic field, and spatial derivatives in $z$.
The radial direction velocity, $u_{r,\alpha}$, can be obtained from the $z$ component of Eq.~\ref{eq:mtm} yielding
\begin{equation}
	u_{r,\alpha} = \frac{1}{n_\alpha q_\alpha B} \frac{\partial P_{zz,\alpha}}{\partial z} - \frac{E_z}{B},
	\label{eq:uy_force_balance}
\end{equation}
which for a Maxwellian distribution, as is the case for the plasma in the center of the domain, becomes
\begin{equation}
	u_{r,\alpha} = \underbrace{\frac{m_\alpha}{n_\alpha q_\alpha B} \frac{ \partial \big( n_\alpha u_{z,\alpha}^2 \big)}{\partial z}}_{\text{inertial}}
	+ \underbrace{\frac{1}{n_\alpha q_\alpha B} \frac{\partial \big( n_\alpha T_{z,\alpha} \big)}{\partial z}}_{\text{diamagnetic}}
	 \underbrace{- \frac{E_z}{B}}_{\mathbf{E}\times\mathbf{B}}.
\end{equation}
Therefore, $u_{r,\alpha}$ is a combination of the inertial, diamagnetic, and $\mathbf{E}\times\mathbf{B}$ drifts.\cite{hainesParticleOrbitsDiamagnetism1978,hainesPhysicsDenseZPinch1982,hainesReviewDenseZpinch2011}
The previous literature\cite{hainesParticleOrbitsDiamagnetism1978,hainesPhysicsDenseZPinch1982,hainesReviewDenseZpinch2011} discusses how the variation in the radial pinch profiles, as in Eqs.~\ref{eq:n_Bennett} and \ref{eq:B_Bennett},\cite{hainesReviewDenseZpinch2011,allenBennettPinchNonrelativistic2018} induce guiding center drifts in the axial direction.
In our work, we find variations in the axial direction, as in Figs.~\ref{fig:potential} and \ref{fig:nu_x}, that will lead to drifts in the radial direction.
Figure~\ref{fig:u_y_force} shows the radial ion (a) and electron (b) velocities for the \SI{10}{\kilo\volt} case in the solid black lines.
The resulting force balanced velocity from Eq.~\ref{eq:uy_force_balance} (dashed blue line) matches the velocity in the center of the domain.
Near the walls, the force balance breaks down, especially in the ions, as the distribution becomes highly non-Maxwellian and the fluid momentum equation, Eq.~\ref{eq:mtm}, becomes invalid.

The individual drifts are also plotted in Fig.~\ref{fig:u_y_force}:
inertial drift (solid green line),
diamagnetic drift (solid purple line),
and $\mathbf{E}\times\mathbf{B}$ drift (solid gray line).
For both species, the diamagnetic and $\mathbf{E}\times\mathbf{B}$ drifts are the dominant terms with the inertial drift playing a negligible role.

The radial ion velocity is entirely negative;
therefore, the ions are moving toward the center of the Z-pinch.
This effect will help in better confinement of the fusion plasma as the ions will converge toward regions of higher density and temperature.

\begin{figure}[!htb]
	\centering
	\includegraphics[width=\linewidth]{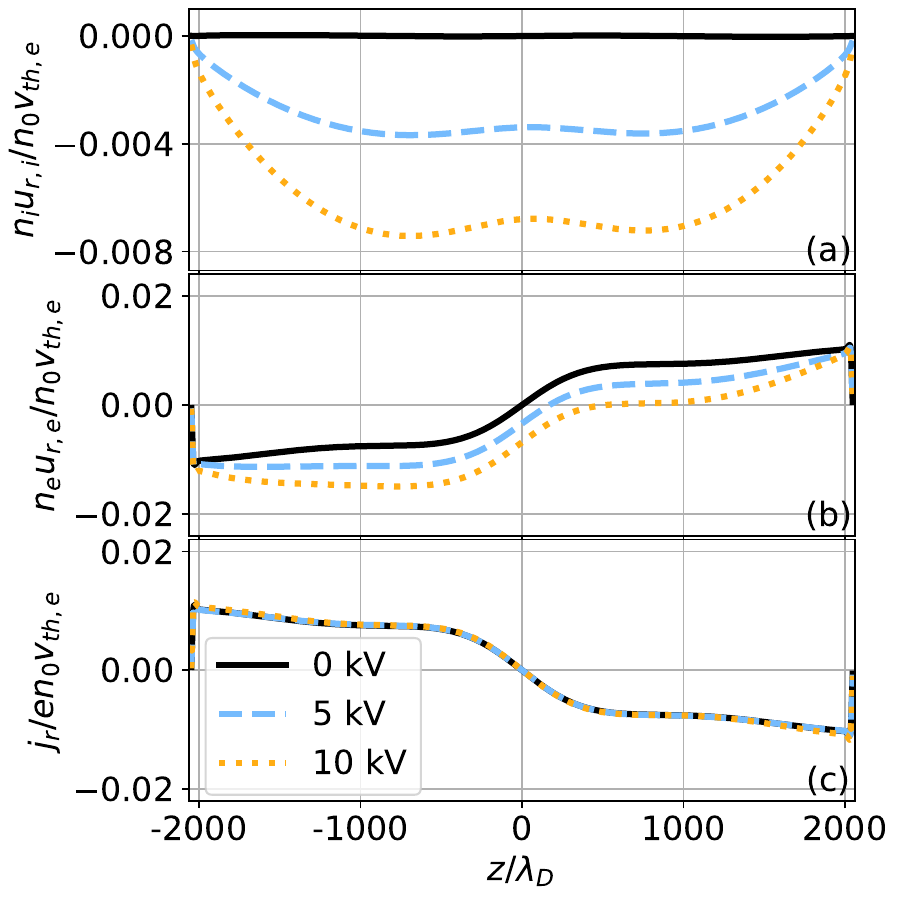}
	\caption{Plots of the radial ion particle flux (a), electron particle flux (b), and current density (c) for the 0 (solid black line), 5 (dashed blue line), and 10 (dotted yellow line) \si{\kilo\volt} cases.
	As bias potential increases, both the radial ion and electron particle fluxes become more negative.
	The radial current density is constant with bias potential in the center of the domain and increases slightly in magnitude near the walls.}
	\label{fig:nu_y}
\end{figure}

In contrast, the electrons move inward closer to the anode and outward closer to the cathode, resulting in a radial shear flow, which is different from the axial shear flow used to stabilize Z-pinches.\cite{shumlakShearedFlowStabilizedZPinch2012,shumlakZpinchFusion2020}
With no bias potential, the point at which $u_{r,e}=0$ is in the center of the domain, as shown in Fig.~\ref{fig:nu_y}(b).
As bias potential increases, this point shifts toward the cathode.
Therefore, at higher bias potentials, the electrons at the pinch radius are traveling toward the center of the Z-pinch for a larger portion of the domain.

The signed charge term, $q_\alpha$, in the diamagnetic drift in Eq.~\ref{eq:uy_force_balance} causes a net current to exist in the radial direction.
Figure~\ref{fig:nu_y} shows the ion particle flux (a), electron particle flux (b), and current density (c) in the radial direction.
The radial ion and electron particle fluxes becomes more negative with increasing bias potential, resulting in larger net flow toward the center of the Z-pinch.

At \SI{0}{\kilo\volt}, the radial ion particle flux mirrors the axial ion particle flux because of the Lorentz force, as noted in previous studies;\cite{moritzPlasmawallTransitionLayers2018}
this effect, however, is very small due to the higher ion mass and is not visible compared to the significantly larger bulk drifts from Eq.~\ref{eq:uy_force_balance} at the higher bias potentials.
As bias potential increases, the radial ion particle flux increases in magnitude with a net flow across the entire domain in the $-r$ direction.
Since both the ion and electron flows increase in the same direction and with similar magnitudes, the current density does not significantly change with bias potential.
The radial current density is symmetric about the origin with positive (radially outward) current near the anode and negative (radially inward) current near the cathode.
Furthermore, the radial current density, which is on the order of \SI{e9}{\ampere \meter^{-2}}, is significantly larger than that in the axial direction, which is only on the order of \SI{e6}{\ampere \meter^{-2}}. This is because the Lorentz force acts more strongly on the electrons due to their smaller mass. Therefore, the three order of magnitude difference in the axial and radial currents is explained by the proton-electron mass ratio, which is 1836.

\section{Summary and Conclusions} \label{s:conclusions}

We perform 1X-2V Boltzmann-Poisson simulations of a proton-electron Z-pinch plasma between two biased electrodes in the presence of a uniform background magnetic field parallel to the walls.
The simulation is an axial cut at the pinch radius ($r=a$), which is the location of peak magnetic field.
To model the larger experiment size in our smaller simulation domain, we artificially increase the collision frequency to improve thermalization of the plasma in the center of the domain.
This has the added effect, however, of artificially increasing the perpendicular mobility and slightly de-magnetizing the ions.

Despite the parallel magnetic field, the artificial collisions de-magnetize the ions sufficiently enough to prevent the formation of an inverse sheath.
Instead, the electrons, which are strongly magnetized, gyrate into the wall first, charging it negative.
This results in a classical sheath near the wall within the electron gyroradius with more ions than electrons, an increasing electric field toward the wall, and the ions accelerating past the Bohm criterion.
However, because the sheath is caused by electron gyrations, the bias potential does not affect the sheath structure, unlike in unmagnetized cases.

In addition, the potential reaches a peak value about one electron gyroradius from the wall before decreasing further in the domain.
This is because the ions further into the domain gyrate into the wall slightly decreasing the ion density creating a non-monotonic potential profile.
Future work should investigate how this behavior changes with collisionality as lower collisions (which are expected in realistic Z-pinch plasmas) may cause an inverse sheath to form instead.\cite{moritzPlasmawallTransitionLayers2018}

Despite the high bias potentials, the current flowing through the plasma in the axial direction is small and driven by changes in the ion motion, unlike in unmagnetized sheaths which develop large currents based on changes in the electron motion.
In our case, the electrons are strongly magnetized and thus do not significantly change with bias potential.
The axial ion particle flux becomes larger with bias potential because the ions are slightly de-magnetized.
Compared to the unmagnetized theoretical\cite{stangebySection26Potential2000} axial current density of \SI{e9}{\ampere\meter^{-2}}, our magnetized simulations only have axial current densities on the order of \SI{e6}{\ampere\meter^{-2}}.
Therefore, there is significantly less current further from the center of the Z-pinch.
This effect is expected to be even stronger in lower collisionality plasmas that have lower perpendicular mobility.

In the radial direction, there are significant flows at higher bias potentials due to the axial force balance between the Lorentz force and pressure tensor.
The radial current density is on the order of \SI{e9}{\ampere\meter^{-2}}, which is three orders of magnitude larger than the axial current density.
The ions move entirely toward the radial center of the Z-pinch improving confinement of the fusion fuel.
The electrons, however, move inward toward the center of the Z-pinch closer to the anode and outward closer to the cathode.
Furthermore, as bias potential increases, the region of inward motion becomes larger as the $u_{r,e}=0$ point moves closer to the cathode.
This resulting radial shear flow is different from the axial shear flow used to stabilize Z-pinches.

The work presented here only examines what happens in the axial cut at $r=a$. 
Future work should not only use realistic collision frequencies, but also examine the transition region from the unmagnetized case at $r=0$ to the fully magnetized case at $r=a$.

\begin{acknowledgments}
	
	The authors acknowledge funding from Dr. Lindsay V. Goodwin's New Jersey Institute of Technology startup funds
	and Dr. Bhuvana Srinivasan's University of Washington startup funds.
	
	The authors acknowledge the Advanced Research Computing Services at New Jersey Institute of Technology for providing computational resouces and technical support that have contributed to the results in this paper.
	
\end{acknowledgments}

\section*{Data Availability Statement}    

Readers may reproduce our results and also use Gkeyll for their applications. The code and input files used here are available online. 
Full installation instructions for Gkeyll are provided on the Gkeyll website at \url{https://gkeyll.readthedocs.io/}. The code can be installed on Unix-like 
operating systems (including Mac OS and Windows using the Windows Subsystem for Linux) either by installing the pre-built binaries using 
the conda package manager (\url{https://www.anaconda.com}) or building the code via sources. The input files used here are under version 
control and can be obtained from the repository at \url{https://github.com/ammarhakim/gkyl-paper-inp/tree/master/2026_PoP_Skolar_Invited}.

\bibliography{reference}

\end{document}